\documentclass[preprint,prd,aps,showpacs,showkeys,nofootinbib]{revtex4-1}
\usepackage{amssymb}
\usepackage{amsmath}
\usepackage{graphicx}
\usepackage{subfigure}
\usepackage{float}
\usepackage{mathrsfs}
\usepackage{amsmath}
\begin{document}

\title{Constraining gravitational-wave polarizations with Taiji}

\author{Chang Liu$^{1,2}$}
\email{liuchang@itp.ac.cn}

\author{Wen-Hong Ruan$^{1,2}$}
\email{ruanwenhong@itp.ac.cn}

\author{Zong-Kuan Guo$^{1,2,3}$}
\email{guozk@itp.ac.cn}

\affiliation{$^1$CAS Key Laboratory of Theoretical Physics, Institute of Theoretical Physics, Chinese Academy of Sciences, P.O. Box 2735, Beijing 100190, China}
\affiliation{$^2$School of Physical Sciences, University of Chinese Academy of Sciences, No.19A Yuquan Road, Beijing 100049, China}

\affiliation{$^3$School of Fundamental Physics and Mathematical Sciences, Hangzhou Institute for Advanced Study,
University of Chinese Academy of Sciences, Hangzhou 310024, China}

\begin{abstract}
Space-based gravitational-wave detectors consist of a triangle of three spacecraft,
which makes it possible to detect polarization modes of gravitational waves
due to the motion of the detectors in space.
In this paper we explore the ability of Taiji to detect the polarization modes
in the parametrized post-Einsteinian framework.
Assuming massive black hole binaries with the total mass of $M=4\times10^5\,M_{\odot}$ at redshift of $z=1$,
we find that Taiji can measure the dipole and quadruple emission ($\Delta\alpha_D/\alpha_D$ and $\Delta\alpha_Q/\alpha_Q$)
with the accuracy of up to $\sim 0.04\%$, 
with the fiducial value of $\alpha_D=0.001$, 
the scalar transverse and longitudinal modes ($\Delta\alpha_B$ and $\Delta\alpha_L$) up to $\sim 0.01$,
and the vector modes ($\Delta\alpha_V$) up to $\sim 0.0005$.
\end{abstract}

\maketitle

\section{Introduction}
So far, general relativity (GR) has passed a large number of experiments
from solar system and binary pulsars~\cite{Stairs:2003eg,Will:2014kxa,Wex:2014nva}.
Recently, the direct detections of gravitational waves (GWs) with advanced LIGO allow us to
test GR in strong-field regime~\cite{Abbott:2016blz,LIGOScientific:2019fpa}.
In GR, GWs contain only two transverse-traceless polarization modes~\cite{Eardley:1974nw}.
However, in alternative theories of gravity, GWs can have up to six polarization modes~\cite{Eardley:1973br,Eardley:1974nw,Will1993}.
For example, in Brans-Dicke theory there exists one scalar polarization mode
in addition to the two transverse-traceless modes of GR~\cite{Brans:1961sx},
while in $f(R)$ gravity or screened modified gravity there are two additional scalar polarization modes~\cite{Alves:2009eg,DeFelice:2010aj,Rizwana:2016qdq,Liang:2017ahj,Niu:2019ywx}.
Einstein-Aether theory predicts the existence of scalar and vector polarization modes~\cite{Jacobson:2004ts,Lin:2018ken,Zhang:2019iim},
while generalized tensor-vector-scalar theories, such as TeVeS theory, predict the existence of all 6 polarization modes~\cite{Bekenstein:2004ne,Gong:2018ybk}.
Therefore, the probe of the additional polarization modes allows us to capture deviations from GR~\cite{Abbott:2018utx}.

However, it is hard to detect the additional polarization modes of GWs from compact binary coalescence signals with advanced LIGO
because its two detectors are co-oriented~\cite{Isi:2017fbj,Takeda:2018uai}.
The future ground-based detector network of advanced LIGO with advanced Virgo, KAGRA, and LIGO India,
has the ability of probing additional polarization modes~\cite{Yunes:2013dva,Hagihara:2019ihn}.
Recently, it is pointed out that a single third-generation ground-based detector could be regarded as a virtual detector network due to a long signal and the Earth's rotation, whose time-varying antenna pattern functions play an important role in testing GW polarizations~\cite{Takeda:2019gwk}. Additional polarizations can also be tested from transient burst GWs~\cite{Hayama:2012au}, continuous GWs~\cite{Isi:2015cva}, and stochastic GWs~\cite{Nishizawa:2009bf,Callister:2017ocg,Abbott:2018utx}.
In addition, the space-based GW detectors, such as LISA~\cite{Audley:2017drz} and Taiji~\cite{Hu:2017mde},
consist of a triangle of three spacecraft in orbit around the Sun,
which make it possible to detect the additional polarization modes of GWs
due to the motion of the detectors in space~\cite{Gair:2012nm,Philippoz:2017}. In our analysis, we focus on massive black hole binaries (MBHBs).

MBHBs are one of the main targets of space-based GW observatories,
which are expected to be detected with extremely high signal-to-noise ratios (SNRs),
so the main purpose of this paper is to investigate the potential constraints on additional polarization modes
of GWs with Taiji using the inspiral phase of MBHBs.
To make our analysis  model-independent, we use the parametrized post-Einsteinian (ppE) formalism,
developed in Refs.~\cite{Yunes:2009ke,Chatziioannou:2012rf} to parametrize the effects on the non-GR polarization modes in modified gravity theories. 
Although the ppE formalism can describe the most well-motivated modified gravity,
such as Brans-Dicke theory, massive gravity, and bimetric theory,
it cannot parametrize all kinds of deviations from GR~\cite{Yunes:2009ke}.
The scope of the application of the ppE formalism is discussed in detail in Ref.~\cite{Yunes:2013dva}.

The paper is organized as follows.
In Section~\ref{sec:wave}, we describe the leading order time-domain ppE waveforms for MBHBs.
In Section~\ref{sec:signal}, the response signal of Taiji is obtained by using the rigid adiabatic approximation.
In Section~\ref{sec:result}, we calculate the root-mean-square errors of the ppE parameters by using the Fisher-matrix method.
The last section is devoted to conclusions.
Throughout this paper, we use units with $G=c=1$, where $G$ is the gravitational constant and $c$ is the speed of light.

\section{Parametrized post-Einsteinian waveform \label{sec:wave}}
In general, GWs have six polarization modes: two transverse-traceless modes [plus ($+$) and cross ($\times$)],
two vector longitudinal modes ($U$ and $V$),
a scalar transverse breathing mode ($B$), and a scalar longitudinal mode ($L$).
The wave tensor with all six polarization modes can be written as
\begin{equation}
\label{eq:wtensor}
\mathbf{H}(t) = h_{+}(t) \boldsymbol{\epsilon}^{+}+h_{\times}(t) \boldsymbol{\epsilon}^{\times}+h_{U}(t) \boldsymbol{\epsilon}^{U}+h_{V}(t) \boldsymbol{\epsilon}^{V}+h_{B}(t) \boldsymbol{\epsilon}^{B}+h_{L}(t) \boldsymbol{\epsilon}^{L} ,
\end{equation}
where $\boldsymbol{\epsilon}^{A}$ ($A=+,\times,U,V,B,L$) are the polarization tensors and $h_{A}$ are the waveforms of the polarization modes.
Following~\cite{Rubbo:2003ap} we work in the heliocentric right-handed orthogonal reference frame $(\hat{x},\hat{y},\hat{z})$.
In such a frame, the Sun is located at the origin of the coordinates,
the $x$-axis is in the direction of vernal equinox,
and the $z$-axis is parallel to the orbital angular momentum of the Earth.
For a GW propagating in the $\hat{k}$ direction,
the bases of the source reference frame can be written as
\begin{eqnarray}
\hat{k} &=& -\sin \theta \cos \phi \,\hat{x} - \sin \theta \sin \phi \,\hat{y} - \cos \theta \,\hat{z} , \nonumber \\
\hat{u} &=& \cos \theta \cos \phi \,\hat{x} + \cos \theta \sin \phi \,\hat{y} - \sin \theta \,\hat{z} , \nonumber \\
\hat{v} &=& \sin \phi \,\hat{x}-\cos \phi \,\hat{y} ,
\end{eqnarray}
where $(\theta, \phi)$ is the sky location of the source on the celestial sphere.
Then the polarization tensors in~\eqref{eq:wtensor} are given by
\begin{eqnarray}
\label{epsilon}
\boldsymbol{\epsilon}^{+} &=& \mathbf{e}^{+} \cos 2\psi -\mathbf{e}^{ \times}\sin 2\psi  , \nonumber \\
\boldsymbol{\epsilon}^{\times} &=& \mathbf{e}^{+}\sin 2\psi+\mathbf{e}^{ \times}\cos 2\psi  , \nonumber \\
\boldsymbol{\epsilon}^{U} &=& \mathbf{e}^{U}\cos \psi-\mathbf{e}^{V}\sin \psi  , \nonumber \\
\boldsymbol{\epsilon}^{V} &=& \mathbf{e}^{U}\sin \psi +\mathbf{e}^{V}\cos \psi  , \nonumber \\
\boldsymbol{\epsilon}^{B} &=& \mathbf{e}^{B} , \nonumber \\
\boldsymbol{\epsilon}^{L} &=& \mathbf{e}^{L} ,
\end{eqnarray}
where $\psi$ is the polarization angle and the six basis tensors are~\cite{Nishizawa:2009bf}
\begin{eqnarray}
\mathbf{e}^{+} &=& \hat{u} \otimes \hat{u}-\hat{v} \otimes \hat{v} , \nonumber \\
\mathbf{e}^{\times} &=& \hat{u} \otimes \hat{v}+\hat{v} \otimes \hat{u} , \nonumber \\
\mathbf{e}^{U} &=& \hat{u} \otimes \hat{k}+\hat{k} \otimes \hat{u} , \nonumber \\
\mathbf{e}^{V} &=& \hat{v} \otimes \hat{k}+\hat{k} \otimes \hat{v} , \nonumber \\
\mathbf{e}^{B} &=& \hat{u} \otimes \hat{u}+\hat{v} \otimes \hat{v} , \nonumber \\
\mathbf{e}^{L} &=& \hat{k} \otimes \hat{k} .
\end{eqnarray}
In Ref.~\cite{Chatziioannou:2012rf}, the extended ppE formalism is proposed to construct model-independent tests of GR, including all GW polarization modes. The amplitude and phase of GWs can be obtained from the metric perturbation and energy evolution, respectively. Following Refs.~\cite{Arun:2012hf,Hansen:2014ewa,OBeirne:2019lwp}, we assume that the tensor modes ($+$, $\times$) are dominated by quadrupole radiation and the four additional polarization modes ($U$, $V$, $B$, $L$) are dominated by dipole radiation.  
In Ref.~\cite{Chatziioannou:2012rf}, it is pointed out that dipole radiation leads to a term of $-1$PN order in the waveform
relative to one from quadrupole radiation.
In our analysis, the inspiral is observed for 60 days before it reaches the innermost stable circular orbit (ISCO).
In the inspiral phase
the contributions from dipole radiation are about ten times larger than those from quadrupole radiation.
Hence, for the four additional polarization modes ($U$, $V$, $B$, $L$),
which arise due to the modifications to GR,
we only consider dipole radiation in the inspiral phase.
However, since GR has passed all current experimental tests,
it is reasonable to assume that quadrupole radiation dominates over dipole radiation in the tensor modes ($+$, $\times$).
So, to leading order the waveforms in the ppE framework can be written as~\cite{OBeirne:2019lwp}
\begin{eqnarray}
\label{h_A}
h_{+} &=& \dfrac{2  \mathcal{M}}{ d_L} \left( \mathcal{M} \omega\right)^{2/3} \left(1+\cos^{2}\iota\right) \cos \left(2 \Phi+2 \Phi_{0}\right) , \nonumber \\
h_{\times} &=& \dfrac{4  \mathcal{M}}{ d_L} \left( \mathcal{M} \omega\right)^{2/3} \cos \iota \sin \left(2 \Phi+2 \Phi_{0}\right) , \nonumber \\
h_{U} &=& \dfrac{\alpha_V  \mathcal{M}}{ d_L} \left( \mathcal{M} \omega\right)^{1/3} \cos \iota \cos \left(\Phi+\Phi_{0}\right) , \nonumber \\
h_{V} &=& \dfrac{\alpha_V  \mathcal{M}}{ d_L} \left( \mathcal{M} \omega\right)^{1/3} \sin \left(\Phi+\Phi_{0}\right) , \nonumber \\
h_{B} &=& \dfrac{\alpha_B  \mathcal{M}}{ d_L} \left( \mathcal{M} \omega\right)^{1/3} \sin \iota \cos \left(\Phi+\Phi_{0}\right) , \nonumber \\
h_{L} &=& \dfrac{\alpha_L  \mathcal{M}}{ d_L} \left( \mathcal{M} \omega\right)^{1/3} \sin \iota \cos \left(\Phi+\Phi_{0}\right) ,
\end{eqnarray}
where $\alpha_V$, $\alpha_B$, $\alpha_L$ are the dimensionless ppE parameters,
$d_L$ is the luminosity distance,
$\mathcal{M} = M \eta^{3/5}$ is the chirp mass with the symmetric mass ratio $\eta=m_1 m_2/M^2$, and the total mass $M=m_1+m_2$,
$\iota$ is the inclination angle,
$\Phi$ is the orbital phase of the binary,
and $\Phi_0$ is the initial orbital phase. 
It is reported in~\cite{Chatziioannou:2012rf} that for quadrupole radiation the inclination dependence in the waveforms 
is $\sin^2\iota$ for the scalar modes,
$\sin\iota$ for the $U$ mode, and $\sin 2\iota$ for the $V$ mode
while for dipole radiation, the inclination dependence is $\sin \iota$ for the scalar modes,
$\cos \iota$ for the $U$ mode, and $1$ for the $V$ mode.
Following~\cite{OBeirne:2019lwp}, we have assumed that quadrupole radiation dominates in the tensor modes ($+$, $\times$)
and dipole radiation dominates in the four additional polarization modes ($U$, $V$, $B$, $L$) in Eq.~\eqref{h_A}.
Furthermore, the phase of dipole radiation is equal to the orbital phase while the phase of quadrupole radiation is equal to twice.
The evolution of orbital angular frequency is given by~\cite{OBeirne:2019lwp}
\begin{equation}
\label{wdt}
\dfrac{d \omega}{d t}=\alpha_D \eta^{2/5} \mathcal{M} \omega^{3}+\alpha_Q \mathcal{M}^{5/3} \omega^{11 / 3} ,
\end{equation}
where $\alpha_D$ and $\alpha_Q$ characterize the dipole and quadrupole contribution to the frequency evolution, respectively.
The form of $\alpha_D \eta^{2/5}\mathcal{M}$ in Eq.~\eqref{wdt} is chosen so that $\alpha_D$ is dimensionless.
In a specific theory the parameters $\alpha_D$ and $\alpha_Q$ may relate to  $\alpha_V, \alpha_B, \alpha_L$~\cite{Chatziioannou:2012rf}. To make our analysis independent of a specific model, we treat the ppE parameters ($\alpha_D, \alpha_Q, \alpha_V, \alpha_B, \alpha_L$) as independent, so that we underestimated the constraints on the ppE parameters. In the GR case, $\alpha_D=\alpha_V=\alpha_B=\alpha_L=0$ and $\alpha_Q=96/5$.
From~\eqref{wdt}, we can see that there is a degeneracy between $\alpha_D$ and $\eta$.
The higher-order waveforms in the ppE framework are required to break such a degeneracy.

Since Eq.~\eqref{wdt} does not have an explicit solution for $\omega(t)$, following Ref.~\cite{OBeirne:2019lwp}, we can get $t(\omega)$
\begin{eqnarray}
\label{eq:time}
t &=& t_{0}+\int_{\omega_{0}}^{\omega(t)} \frac{d \omega}{\alpha_D \eta^{2/5}  \mathcal{M} \omega^{3}+\alpha_Q \mathcal{M}^{5/3} \omega^{11 / 3}} \nonumber \\
 &=& t_{0}-\frac{(\omega^{-2}-\omega_0^{-2})}{2  \mathcal{M} \eta ^{2/5} \alpha _D }-\frac{3 \alpha _Q^2 \mathcal{M}^{1/3}}{2 \eta ^{6/5} \alpha _D^3 }\left(\frac{1}{\omega^{2/3}}-\frac{1}{\omega_0^{2/3}}\right)+\frac{3 \alpha _Q \mathcal{M}^{-1/3}}{4 \eta ^{4/5} \alpha _D^2 }\left(\frac{1}{\omega^{4/3}}-\frac{1}{\omega_0^{4/3}}\right) \nonumber \\
&& +\,\frac{3\mathcal{M} \alpha _Q^3}{2 \eta ^{8/5} \alpha _D^4} \log \left(\frac{\eta ^{2/5}\omega^{-2/3}\alpha _D  +   \mathcal{M}^{2/3}\alpha _Q }{\eta ^{2/5}\omega_0^{-2/3}\alpha _D + \mathcal{M}^{2/3}\alpha _Q}\right) ,
\end{eqnarray}
where $\omega_0=\omega(t_0)$ is the initial orbital angular frequency.
The orbital phase is
\begin{eqnarray}
\label{phase}
\Phi(t) &=&\Phi_{0}+\int_{t_{0}}^{t} \omega(t) d t \nonumber \\
&=&\Phi_{0}+\int_{\omega_{0}}^{\omega(t)} \frac{\omega d \omega}{\alpha_D \eta^{2/5}  \mathcal{M} \omega^{3}+\alpha_Q  \mathcal{M}^{5/3} \omega^{11 / 3}}  \nonumber \\
&=&\Phi_{0}-\frac{(\omega^{-1}-\omega_0^{-1})}{\alpha_D \eta ^{2/5}  \mathcal{M}}+\frac{3 \alpha_Q\mathcal{M}^{-1/3}}{\alpha_D^2 \eta ^{4/5}}\left(\frac{1}{\omega^{1/3}}-\frac{1}{\omega_0^{1/3}}\right)\nonumber \\
& &+\frac{3 \alpha_Q^{3/2}}{\alpha_D^{5/2} \eta }\left(\tan ^{-1}\left(\frac{\sqrt{\alpha_Q} (\mathcal{M} \omega)^{1/3}}{\sqrt{\alpha_D} \eta^{1/5}}\right)-\tan ^{-1}\left(\frac{\sqrt{\alpha_Q} (\mathcal{M} \omega_0)^{1/3}}{\sqrt{\alpha_D} \eta^{1/5}}\right)\right) ,
\end{eqnarray}
where $\Phi_{0}=\Phi(t_0)$ is the initial orbital phase~\cite{OBeirne:2019lwp}.

\section{Method \label{sec:signal}}
For space-based GW detectors such as LISA and Taiji,
the motion of the spacecraft in orbit around the Sun will introduce multiple modulations on the GW signals.
In what follows we focus on Taiji.
Under the rigid adiabatic approximation~\cite{Rubbo:2003ap},
the Michelson output with the spacecraft 1 as a synthesized detector can be written as
\begin{equation}
\label{s1t}
h(t) =\Re(\mathbf{F}(t, f(\xi)): \mathbf{H}(\xi)) ,
\end{equation}
where $\Re$ denotes the real part, $\mathbf{a}:\mathbf{b}=a^{\mu\nu}b_{\mu\nu}$, and
\begin{eqnarray}
\mathbf{F}(t,f(\xi))=\frac{1}{2}\left[(\hat{r}_{12}(t) \otimes \hat{r}_{12}(t)) \mathcal{T}(\hat{r}_{12}(t), f(\xi))
 -(\hat{r}_{13}(t) \otimes \hat{r}_{13}(t)) \mathcal{T}(\hat{r}_{13}(t), f(\xi))\right].
\end{eqnarray}
The transfer function is~\cite{Rubbo:2003ap}
\begin{eqnarray}
\label{trans}
\mathcal{T}(\hat{r}_{ij}(t), f(\xi))&=& \frac{1}{2}\left[\operatorname{sinc}\left(\frac{f(\xi)}{2 f_{*}}(1-\hat{r}_{ij}(t) \cdot \hat{k})\right)\right.
 \exp \left(-i \frac{f(\xi)}{2 f_{*}}(3+\hat{r}_{ij}(t) \cdot \hat{k})\right) \nonumber\\
& &+\operatorname{sinc}\left(\frac{f(\xi)}{2 f_{*}}(1+\hat{r}_{ij}(t) \cdot \hat{k})\right)
 \left. \exp \left(-i \frac{f(\xi)}{2 f_{*}}(1+\hat{r}_{ij}(t) \cdot \hat{k})\right)\right],
\end{eqnarray}
where $\operatorname{sinc}(x)\equiv\sin(x)/x$, $\xi(t)=t-\hat{k} \cdot \mathbf{x}(t)$,
and $f_{*}=1/(2 \pi L)$ with the arm-length of Taiji $L=3\times10^9$ m.
The coordinates $\mathbf{x}(t)$ of the three spacecraft in the heliocentric reference frame is given by~\cite{Rubbo:2003ap}
\begin{eqnarray}
x(t) &=& R \cos \alpha+\frac{1}{2} e R\left[\cos (2 \alpha-\beta)-3 \cos \beta\right] , \nonumber\\
y(t) &=& R \sin \alpha+\frac{1}{2} e R\left[\sin (2 \alpha-\beta)-3 \sin \beta\right] , \nonumber\\
z(t) &=& -\sqrt{3} \,e R \cos (\alpha-\beta) ,
\end{eqnarray}
where $R=1$ AU, $e=L/(2\sqrt{3}R)$ is the orbit eccentricity,
$\alpha=2\pi f_m t+\kappa$ with $f_m=1/$year,
and $\beta=2\pi n/3+\lambda $ $ (n=0,1,2)$.
Here $\kappa$ and $\lambda$ are the initial ecliptic longitude and orientation of the spacecraft, respectively.
The direction from the spacecraft $i$ to the spacecraft $j$ is described by
\begin{equation}
\hat{r}_{i j}\left(t\right)=\frac{\mathbf{x}_{j}(t)-\mathbf{x}_{i}(t)}{L}.
\end{equation}

With the time-domain signal $h(t)$, we can get the Fourier transform of the signal
by using the stationary phase approximation~\cite{Yunes:2009yz}.
The GW parameters in the ppE framework are
\begin{equation}
\boldsymbol{\lambda} = \left\{t_c, \Phi_{0}, \theta, \phi, \psi, \iota, \mathcal{M}, d_L, \alpha_D, \alpha_Q, \alpha_V, \alpha_B, \alpha_L\right\} .
\end{equation}
Such a parameter set is different from the one in Ref.~\cite{OBeirne:2019lwp}, in which the parameters are dimensionful.

We use the Fisher matrix method to explore the ability of Taiji to detect deviations from GR.
The method is based on computing the inverse of the Fisher matrix, known as the variance-covariance matrix.
The diagonal elements of the variance-covariance matrix are maximum likelihood estimators
of the variance of parameters around the true value in the case of a large SNR~\cite{Vallisneri:2007ev}.

The Fisher information matrix is defined as
\begin{equation}
\Gamma_{i j} \equiv\left(\frac{\partial h}{\partial \lambda_{i}}, \frac{\partial h}{\partial \lambda_{j}}\right),
\end{equation}
where the noise-weighted inner product is
\begin{equation}
(a, b) = 2 \int_{0}^{\infty} df \frac{\tilde{a}(f) \tilde{b}^{*}(f)+\tilde{a}^{*}(f) \tilde{b}(f)}{S_{n}(f)}.
\end{equation}
Here $S_n(f)$ is the noise power spectral density of Taiji~\cite{Guo:2018npi,Ruan:2019tje,Ruan:2020smc}.
If the noise is Gaussian and stationary, the root-mean-square error in $\lambda_i$ is
\begin{equation}
\Delta \lambda_i = \sqrt{(\Gamma^{-1})_{ii}} ,
\end{equation}
where $\Gamma^{-1}$ is the inverse of the Fisher matrix.
In our analysis, we use two Michelson channels and the combined Fisher matrix is a sum of two Fisher matrices.

\section{Results \label{sec:result}}
To investigate the ability of Taiji to detect the additional polarization modes,
we focus on the parameter estimation of the five ppE parameters ($\alpha_D, \alpha_Q, \alpha_B, \alpha_V, \alpha_L$)
with the fiducial values of $\alpha_D=0.001$,\footnote{
Although $\alpha_D$ vanishes in GR, in our analysis $\alpha_D=0.001$ is chosen to avoid the divergence of \eqref{eq:time} and \eqref{phase}.}
$\alpha_Q=19.2$, $\alpha_V=0$, $\alpha_B=0$, and $\alpha_L=0$.
We consider equal-mass MBHBs with the total mass of $M=4\times10^5\,M_{\odot}$ in the source frame at redshift of $z=1$.
The corresponding luminosity distance can be calculated
in a spatially-flat $\Lambda$CDM Universe with the matter density parameter $\Omega_m=0.3$ and Hubble constant $H_0=67$ km s$^{-1}$ Mpc$^{-1}$.
We consider only the inspiral phase of MBHBs, neglecting all information coming from the merger and the ringdown phases.
In our analysis the coalescence time is chosen to be 60 days
which means the inspiral is observed for 60 days before it reaches the ISCO. 
The upper cutoff frequency is the ISCO frequency $(6^{3/2} \pi M)^{-1}=10.99$ mHz for $M=4\times10^5\,M_{\odot}$. The value of the lower frequency is $0.2427$ mHz, which can be obtained by solving Eq.~\eqref{eq:time}.

\subsection{Inclination angle \label{subsec:ia}}
\begin{figure}[t!]
\includegraphics[width=0.48\textwidth ]{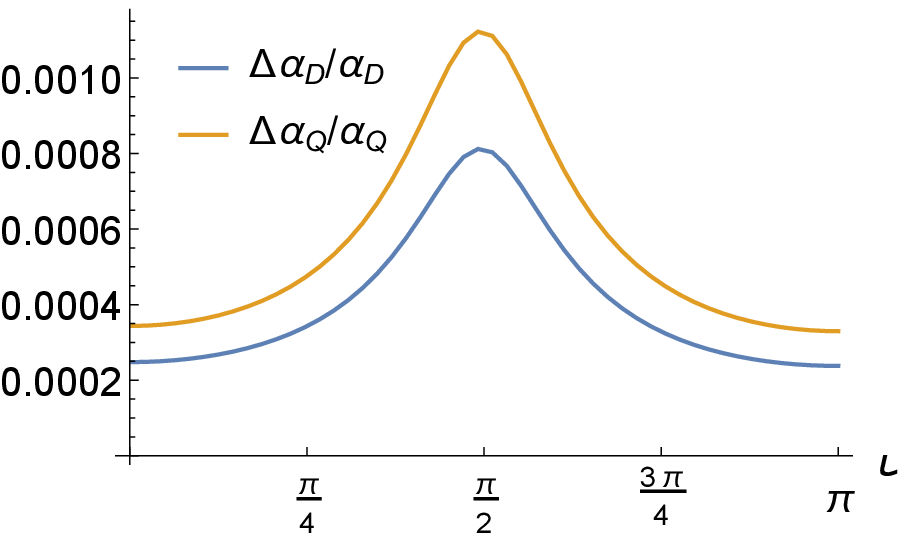}
\includegraphics[width=0.48\textwidth ]{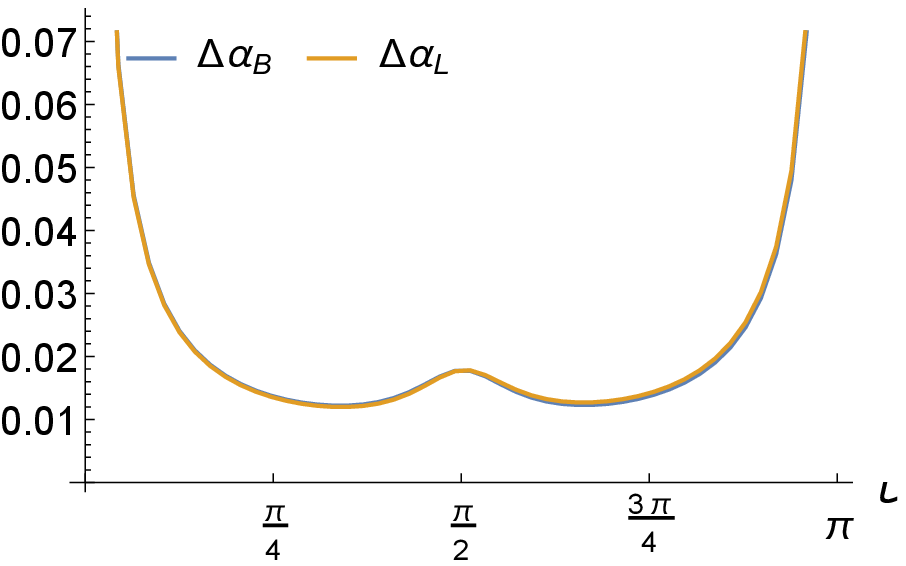}
\includegraphics[width=0.48\textwidth ]{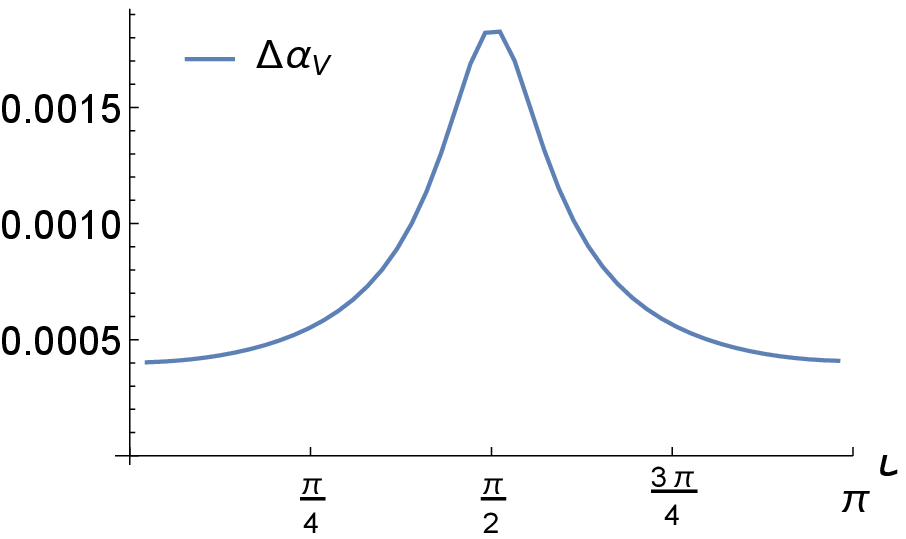}
\includegraphics[width=0.48\textwidth ]{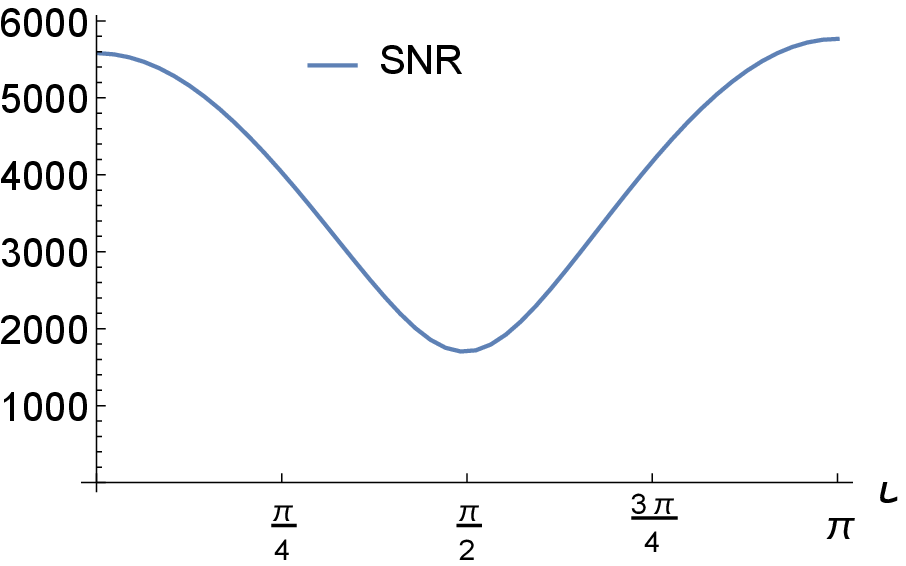}
\caption{Errors in the five ppE parameters and SNRs as a function of the inclination angle
 for equal-mass MBHBs with the total mass of $M=4\times 10^5\,M_{\odot}$ at redshift of $z=1$.}
\label{iota41254pp}
\end{figure}

Setting $\theta=\pi/4$, $\phi=\pi$, and $\psi=0.1$,
we generate $50$ equal-mass MBHBs with the inclination angle from $\iota=0$ to $\pi$.
Figure~\ref{iota41254pp} shows the errors in the five ppE parameters and SNRs as a function of the inclination angle.

From Fig.~\ref{iota41254pp} we see that the measurements of $\alpha_D$, $\alpha_Q$, and $\alpha_V$
are better when $\iota=0$ or $\pi$ (face-on) with larger SNRs than the case of $\iota=\pi/2$ (edge-on).
However, the measurement errors in $\alpha_B$ and $\alpha_L$ increase when the inclination angle approaches $0$ or $\pi$.
This is because 
the $\sin\iota$ dependence of the waveforms in Eq.~\eqref{h_A} implies that
for the scalar modes ($B$ and $L$), the signal for the edge-on binary is stronger than the face-on binary.
Actually, in the heliocentric frame, the signals contain the effects of the relative orientation and motion of
the sources and the detectors by the response functions.
Therefore, the errors in $\alpha_B$ and $\alpha_L$ have a local maximum at $\iota\sim\pi/2$,
as shown in the top-right panel of Fig.~\ref{iota41254pp}.
Our further calculations indicate that the local maximum disappears for sources with some special sky locations.
Since the fiducial values of the five ppE parameters are chosen in GR with only the $+$ and $\times$ modes,
as expected, the SNR for the face-on binary is larger than the edge-on one.
With Taiji, $\Delta\alpha_D/\alpha_D$ and $\Delta\alpha_Q/\alpha_Q$ can be measured with the accuracy of up to $\sim 0.04\%$,
$\Delta\alpha_B$ and $\Delta\alpha_L$ up to $\sim 0.01$, and $\Delta\alpha_V$ up to $\sim 0.0005$.

\subsection{Polarization angle \label{subsec:pa}}
\begin{figure}[t!]
\includegraphics[width=0.48\textwidth ]{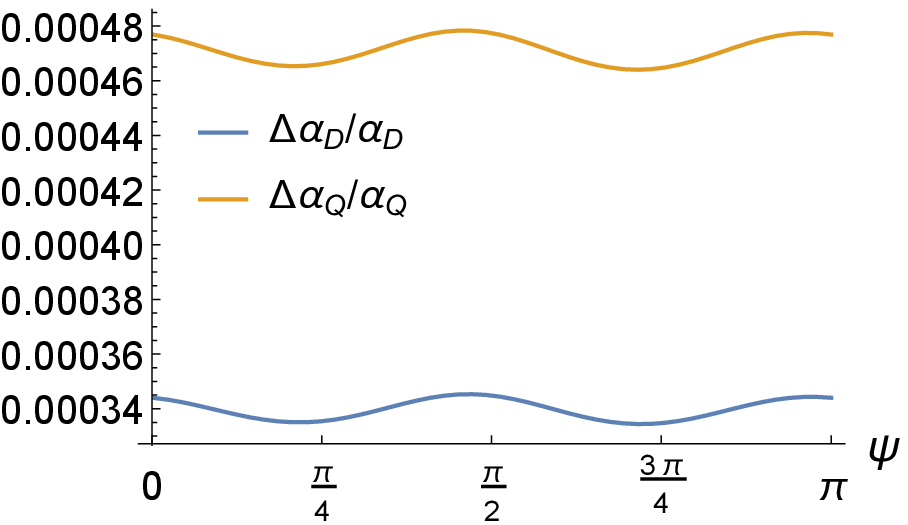}
\includegraphics[width=0.48\textwidth ]{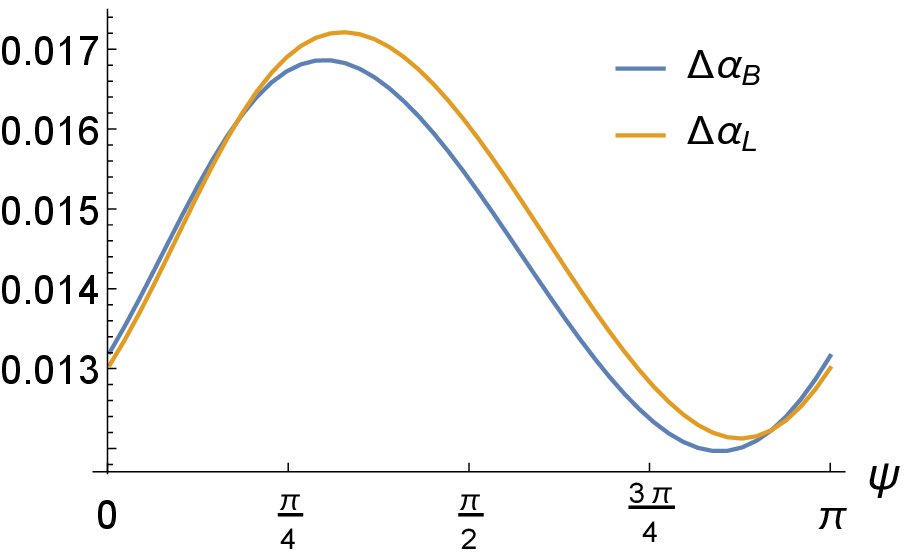}
\includegraphics[width=0.48\textwidth ]{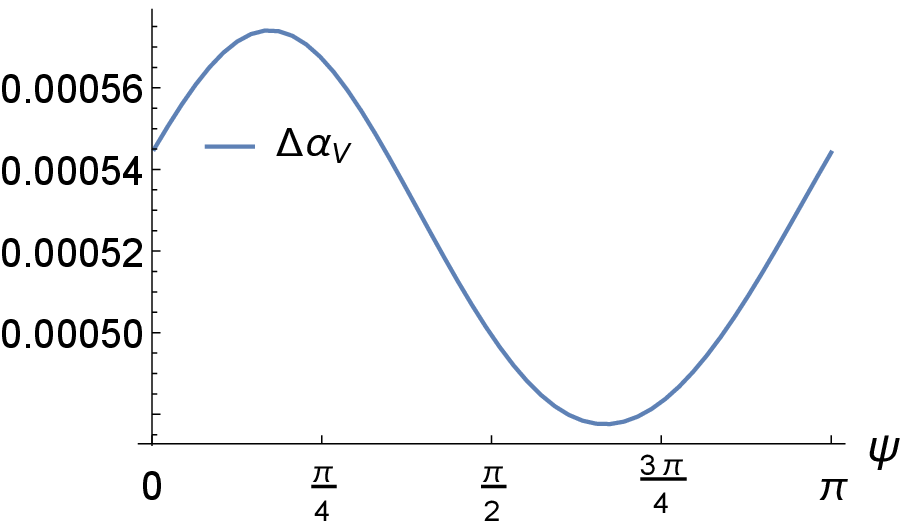}
\includegraphics[width=0.48\textwidth ]{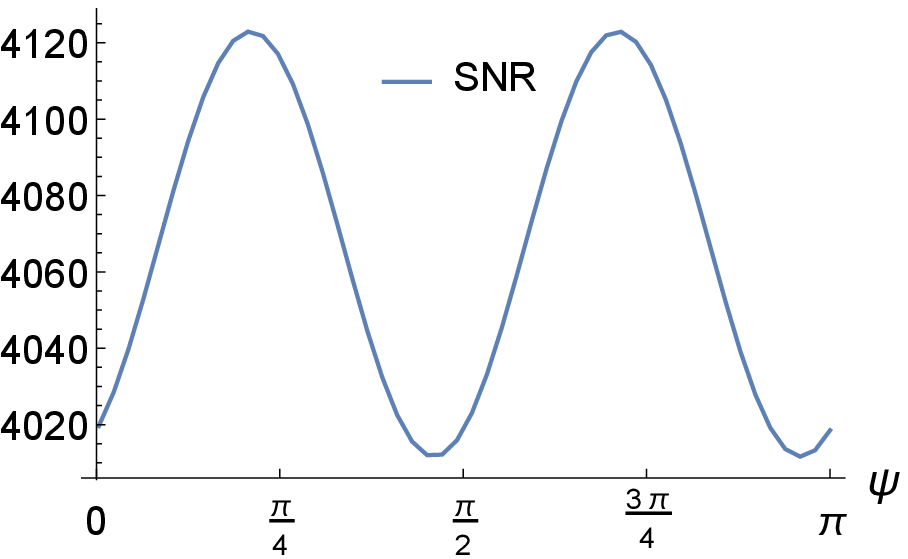}
\caption{Errors in the five ppE parameters and SNRs as a function of the polarization angle.}
\label{psi41254pp}
\end{figure}

We generate $50$ equal-mass MBHB sources in the range of $0 < \psi < \pi$,
with the fixed values of $\theta=\pi/4$, $\phi=\pi$, and $\iota=\pi/4$.
Figure~\ref{psi41254pp} shows the errors in the five ppE parameters and SNRs as a function of the polarization angle.

The $\psi$ dependence of the polarization tensors in Eq.~\eqref{epsilon} indicates that for the $+$ and $\times$ modes the signal has a period of $\pi/2$ while for vector modes the period is $\pi$. In our fiducial values of the ppE parameters, the SNR is determined by the $+$ and $\times$ modes. This is why the SNRs vary with a period of $\pi/2$ and the errors in $\alpha_V$ vary with a period of $\pi$ in Fig.~\ref{psi41254pp}. The errors in $\alpha_D$, $\alpha_Q$ are small for large SNRs with the same period. 
The top-right panel of Fig.~\ref{psi41254pp} shows that the errors in $\alpha_B$ and $\alpha_L$ vary with a period of $\pi$.
Unlike the other modes,
the scalar modes do not explicitly depend on the polarization angle as we can see in Eq.~\eqref{epsilon}.
The errors in $\alpha_B$ and $\alpha_L$ depend on the polarization angle via correlations among other parameters, which depend on $\psi$.

\subsection{Sky location}
\begin{figure}[t!]
\includegraphics[width=0.46\textwidth ]{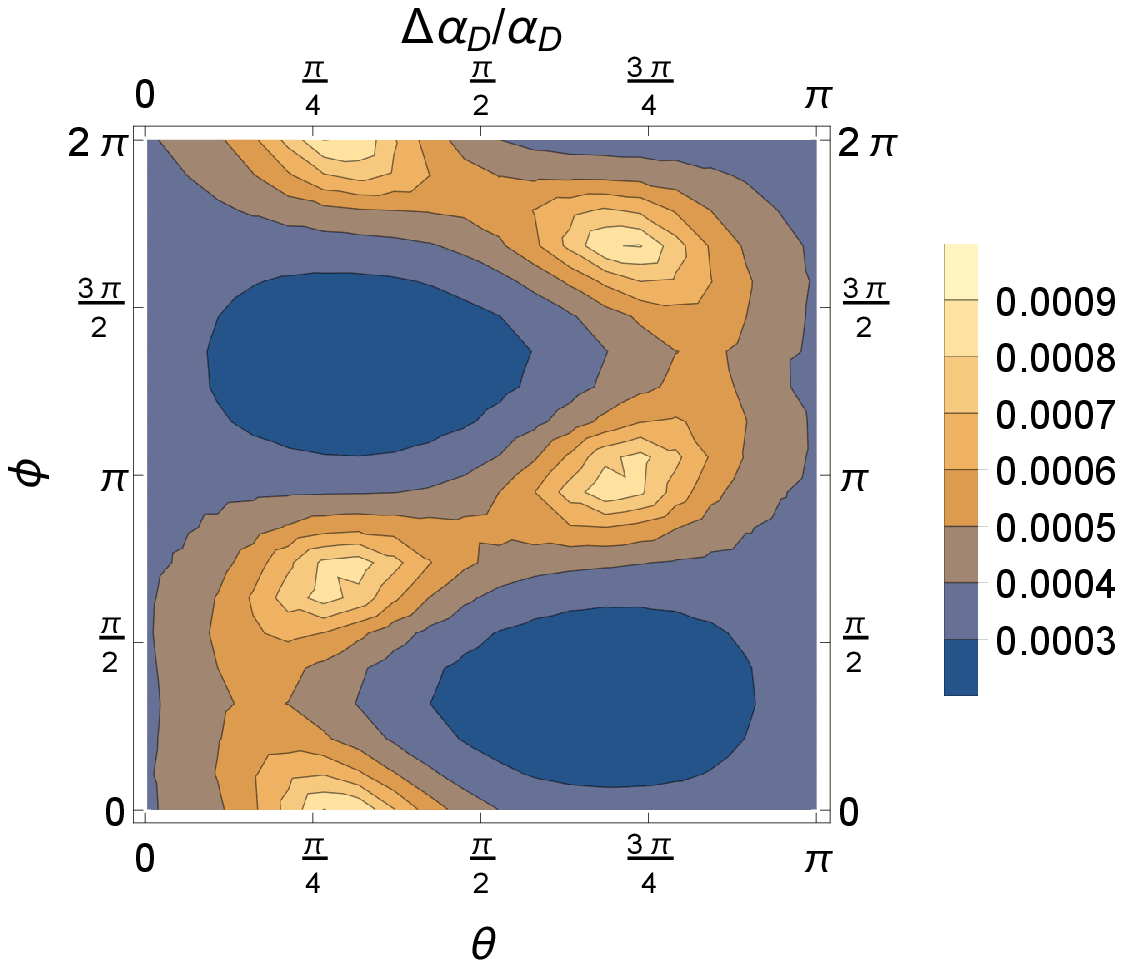}
\includegraphics[width=0.46\textwidth ]{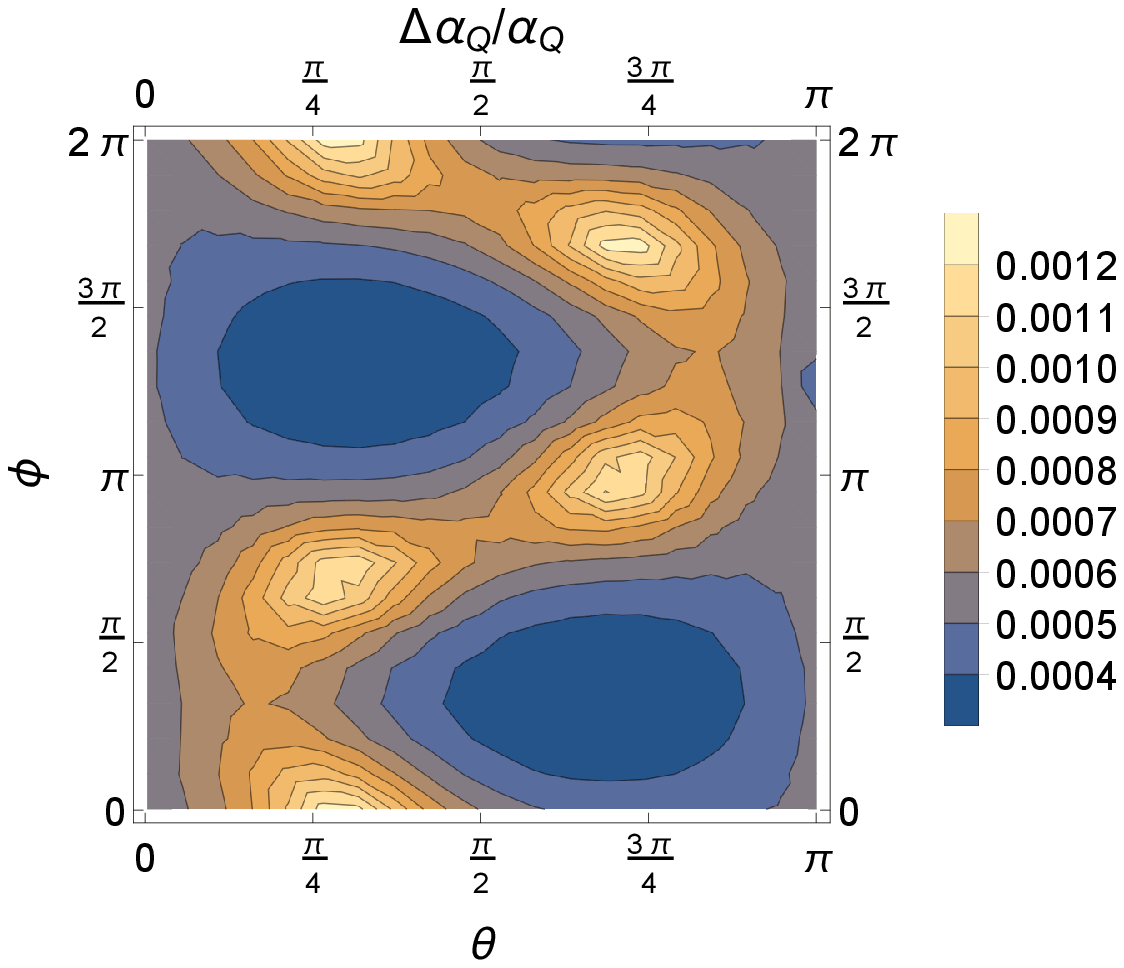}
\includegraphics[width=0.46\textwidth ]{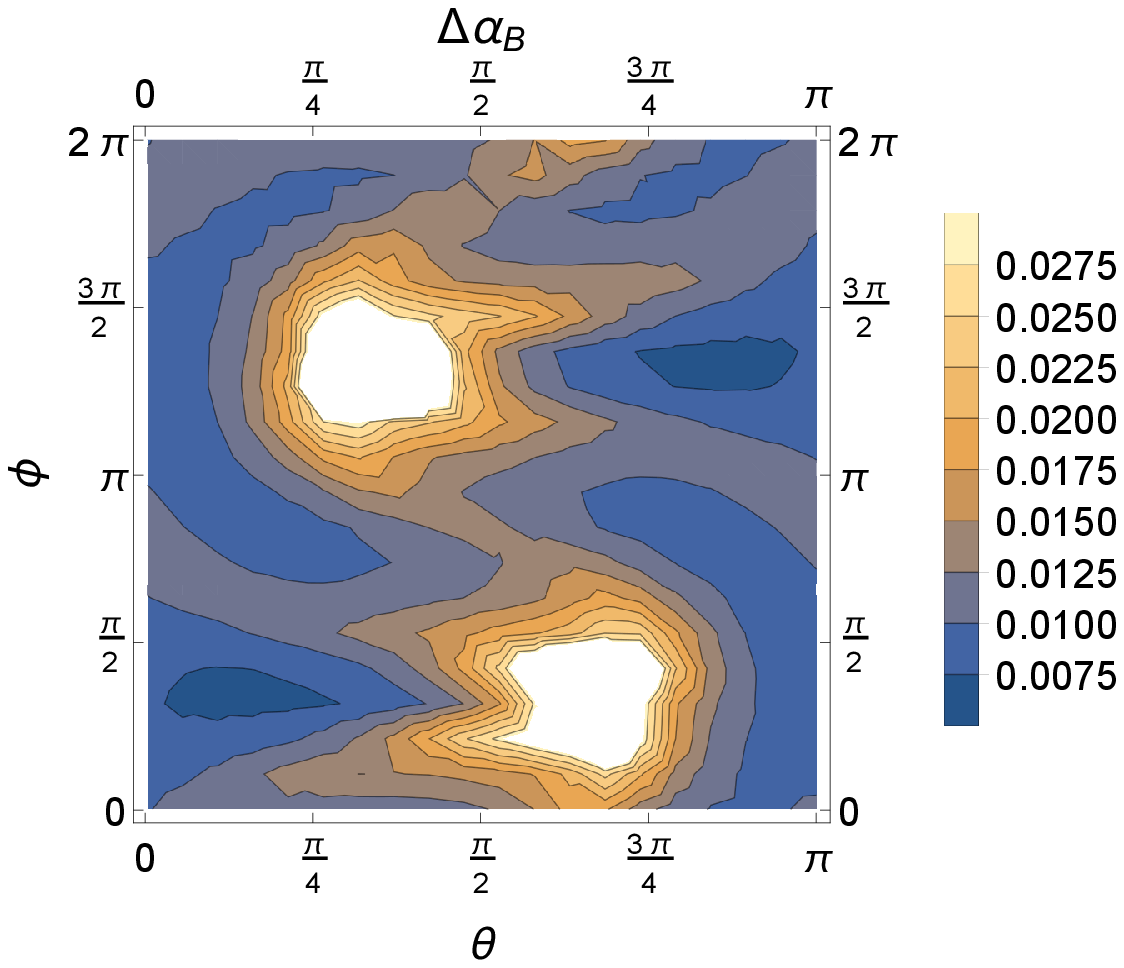}
\includegraphics[width=0.46\textwidth ]{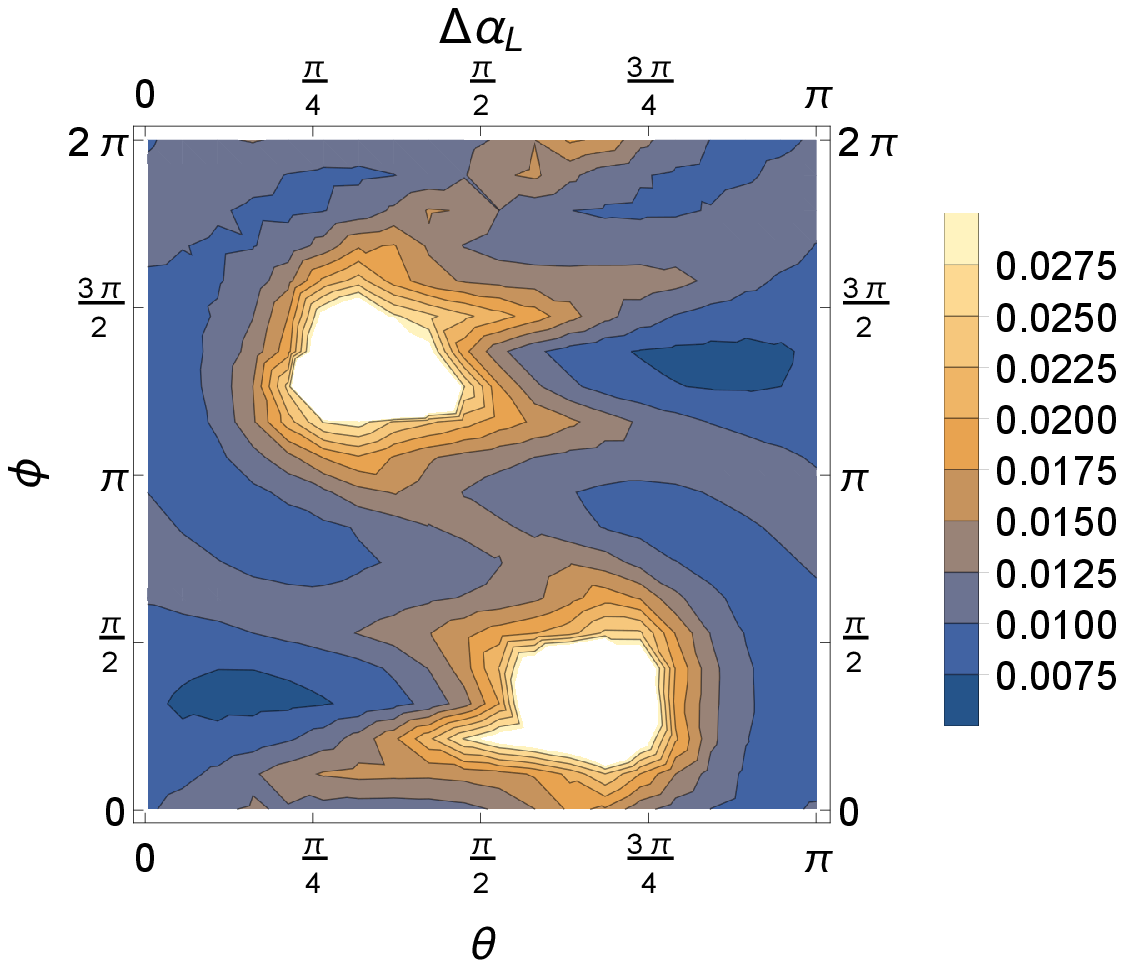}
\includegraphics[width=0.46\textwidth ]{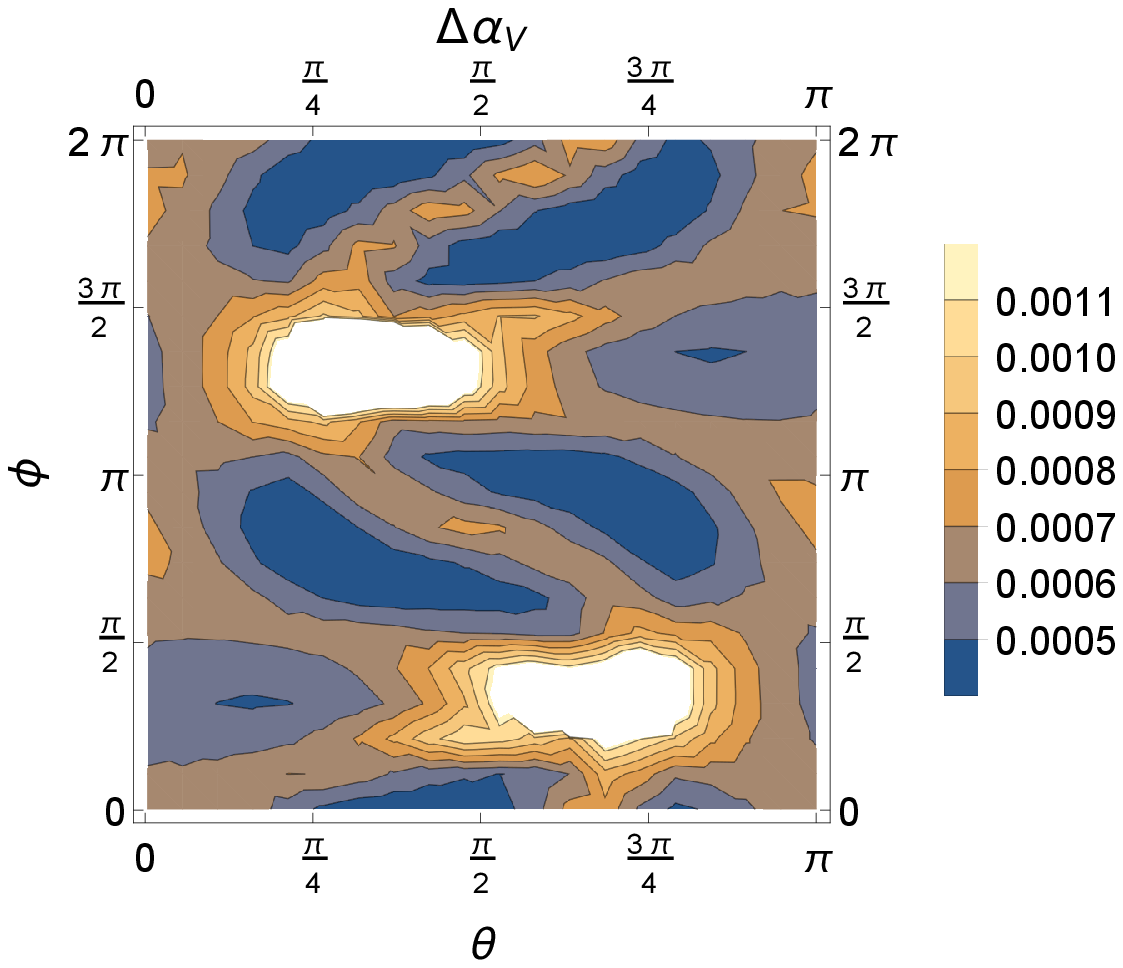}
\includegraphics[width=0.46\textwidth ]{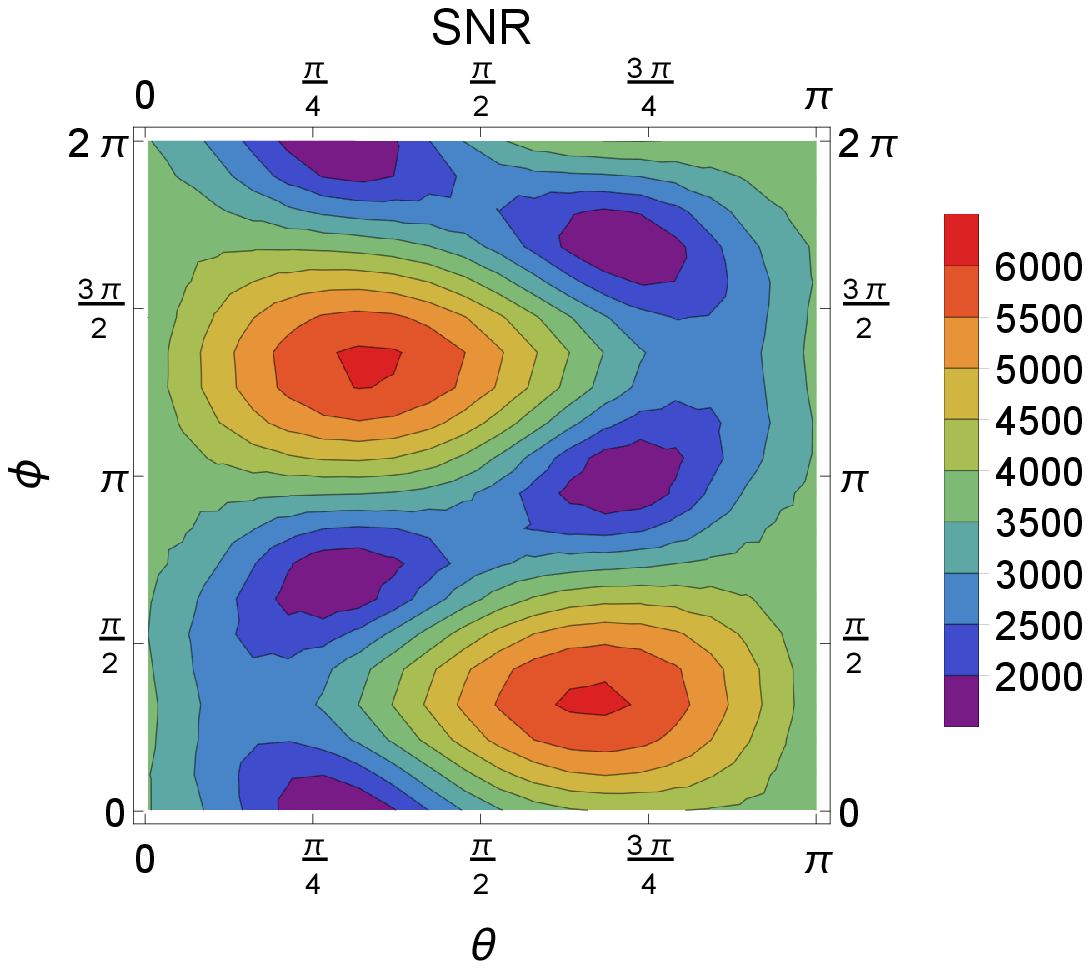}
\caption{Sky-location dependence of the errors in the five ppE parameters and SNRs.}
\label{thephi4125}
\end{figure}

Assuming $\iota=\pi/4$ and $\psi=0.1$,
we generate $400$ equal-mass MBHB sources with different sky locations ($0<\theta<\pi$, $0<\phi<2\pi$)
to illustrate the sky-location dependence of the errors in the five ppE parameters and SNRs in Fig.~\ref{thephi4125}.

We find the errors in $\alpha_D$ and $\alpha_Q$ become small for large SNRs as a function of the sky location.
This is the case as a function of the inclination angle in Sec.~\ref{subsec:ia} and
as a function of the polarization angle in Sec.~\ref{subsec:pa}.
However, for the other three ppE parameters, $\alpha_B$, $\alpha_L$, and $\alpha_V$,
the parameter errors become large for large SNRs as a function of the sky location.
Based on the analysis of the ground-based detector's response for the different polarization modes in Refs.~\cite{Yunes:2013dva,Nishizawa:2009bf}, the detector is insensitive to the vector modes and scalar modes in the direction sensitive to the $+$ and $\times$ modes, and vice versa. Although the response of the space-based detector includes the relative motion between the detector and the source, this result is still valid as shown in middle panels and bottom panels of Fig.~\ref{thephi4125}, so in the direction with large SNRs, determined by $+$ and $\times$ modes, the errors in $\alpha_B$, $\alpha_L$, and $\alpha_V$ are large.
Since the fiducial values of the ppE parameters are chosen in GR,
as expected, the sky-location dependence of the SNRs is consistent with the result of LISA~\cite{Cornish:2018dyw}.
Note that the parameter errors and SNRs are predominantly axisymmetric.

\subsection{Fiducial values}

Now, we study the effects of the fiducial values on our results. We choose the fiducial values of $\alpha_D$, $\alpha_Q-19.2$, $\alpha_B$, $\alpha_L$, and $\alpha_V$ in the range from $10^{-5}$ to $10^{-3}$. 
For $\alpha_Q$, $\alpha_B$, $\alpha_L$, and $\alpha_V$, the errors in the five ppE parameters are almost unchanged. For $\alpha_D$, the errors in $\alpha_B$, $\alpha_L$, and $\alpha_V$ are almost unchanged, but the errors in $\alpha_D$ and $\alpha_Q$ depend on the fiducial value of $\alpha_D$.
Choosing the fiducial values of $\alpha_D=10^{-5}$, $10^{-4}$, and $10^{-3}$,
we obtain $\Delta\alpha_D/\alpha_D\approx 7.10\times10^{-12}$, $1.09\times10^{-7}$, and $3.42\times10^{-4}$; 
and $\Delta\alpha_Q/\alpha_Q\approx 1.19\times10^{-11}$, $8.60\times10^{-8}$, and $4.75\times10^{-4}$, respectively.
This implies that the estimated errors in $\alpha_D$ and $\alpha_Q$ depend strongly on the fiducial value of $\alpha_D$. The errors in $\alpha_D$ and $\alpha_Q$ become small for small fiducial values of $\alpha_D$.
This is because $\alpha_D$ appears in the denominator in the Fisher matrices due to the derivative of the phase function \eqref{phase} with respect to $\alpha_D$ and $\alpha_Q$, which makes the results sensitive to the value of $\alpha_D$. Therefore, the value of $\alpha_D$ affects the measurement errors in $\alpha_D$ and $\alpha_Q$, but the values of $\alpha_Q$, $\alpha_B$, $\alpha_L$, and $\alpha_V$ do not affect the measurement errors.

\section{Conclusions}
We have investigated the ability of Taiji to detect the additional polarization modes of GWs
by using the ppE waveforms for the inspiral phase of MBHBs.
The Fisher matrix method is used to compute the parameter errors in the five ppE parameters,
$\alpha_D$, $\alpha_Q$, $\alpha_B$, $\alpha_V$, and $\alpha_L$.
In our analysis, the fiducial values of the ppE parameters are set in the GR case.
The coalescence time is chosen to be 60 days
for equal-mass MBHBs with the total mass of $M=4\times10^5\,M_{\odot}$ at redshift of $z=1$.
Moreover, we have studied the behavior of the parameter errors
as functions of the inclination angle, polarization angle, and direction to the binary, respectively.

The behavior of $\Delta\alpha_D/\alpha_D$ is the same as that of $\Delta\alpha_Q/\alpha_Q$
as functions of $\iota$, $\psi$, and the sky location.
Both $\Delta\alpha_D/\alpha_D$ and $\Delta\alpha_Q/\alpha_Q$ become small for large SNRs,
which can be measured with the accuracy of up to $\sim 0.04\%$, 
with the fiducial value of $\alpha_D=0.001$.
Although the waveforms of the scalar transverse and longitudinal modes in the ppE framework are the same in Eq.~\eqref{h_A},
the polarization tensors and response functions help us to break their degeneracy. 
However, for the ground-based detector,
the arm length is much smaller than the wavelength of GWs [$f\ll f_{*}=1/(2\pi L)$].
In this case, the transfer function $\mathcal{T}[\hat{r}_{ij}(t), f(\xi)] \approx 1$~\cite{Rubbo:2003ap}.
Due to such a trivial transfer function the breathing and longitudinal scalar modes are completely degenerated~\cite{Nishizawa:2009bf}.
Compared to the ground-based detector,
the transfer function of the space-based detector contains much information
which can break the degeneracy.
For Taiji, the transfer frequency $f_*=15.9$ mHz.
In our analysis, we only consider the inspiral phase of MBHBs with the total mass of $M=4\times10^5\,M_{\odot}$,
choosing the upper cutoff frequency of $10.99$ mHz.
This is the reason why the patterns are very similar between the breathing and longitudinal modes in the middle panels of Fig.~\ref{thephi4125}.
In real data analysis,
the addition of the merger and ringdown phases
will significantly improve measurements of the scalar polarization modes.
Therefore the behavior of $\Delta\alpha_B$ is similar to that of $\Delta\alpha_L$
as functions of the inclination angle, polarization angle, and sky location.
The measurement accuracy of up to $0.01$ can be achieved.
Compared to the scalar polarization modes,
GWs in the inspiral phase of MBHBs are sensitive to the vector polarization modes,
which is measured with the accuracy of up to $0.0005$.

We do not know the true values of the ppE parameters in advance. In fact, we obtained the measurement accuracy of the ppE parameters with Taiji. If the values of the ppE parameters predicted by a specific modified theory of gravity are less than the errors derived in this paper, we can say Taiji does not rule out the theory. If the predicted values are larger than the estimation errors, the additional polarization modes are expected to be detected by Taiji.
Here we emphasize that this is not the case for $\alpha_D$ and $\alpha_Q$
due to the dependence of $\Delta\alpha_D$ and $\Delta\alpha_Q$ on the fiducial value of $\alpha_D$.

\begin{acknowledgements}
This work is supported in part by the National Natural Science Foundation of China Grants No.12075297, No. 11690021, and No. 11851302,
in part by the Strategic Priority Research Program of the Chinese Academy of Sciences Grants No. XDB23030100 and No. XDA15020701,
and by Key Research Program of Frontier Sciences, CAS.
\end{acknowledgements}


\end{document}